\documentclass{article}

\title{Exact Solution of the Klein-Gordon Equation for the Hydrogen Atom Including Electron Spin}
\author{Robert Ducharme}

\begin{document}
\maketitle

\centerline{151 Fairhills Dr., Ypsilanti, MI 48197}
\centerline{E-mail: ducharme01@comcast.net}

\begin{abstract}
The term describing the coupling between total angular momentum and energy-momentum in the hydrogen atom is isolated from the radial Dirac equation and used to replace the corresponding orbital angular momentum coupling term in the radial K-G equation. The resulting spin-corrected K-G equation is a second order differential equation that contains no matrices. It is solved here to generate the same energy eigenvalues for the hydrogen atom as the Dirac equation. 
\end{abstract}

\section{Introduction}
The K-G equation is the quantum mechanical expression of the relativistic energy-momentum relationship. Naturally, the principle of energy-momentum conservation is applicable to all particles, irrespective of their spin, but the spin is relevant in so far as it affects the energy-momentum of the particle. The purpose of this paper is to derive a K-G equation that describes the hydrogen atom by including the affect of the spin on energy-momentum in the Hamiltonian for the electron.

The more conventional approach to the hydrogen atom is to factorize the free-field K-G equation using Dirac matrices giving the Dirac equation. Electromagnetic interactions are then introduced into the Hamiltonian of the Dirac equation through minimal coupling. The result is a system of first-order differential equations that operate on the components of a bi-spinor wavefunction.

The spinless K-G equation for the hydrogen atom is introduced in section 2 of this paper. A key point of interest is that the angular momentum term in the radial component of this equation can be expressed in terms of a simple function $\eta_{\epsilon \kappa}$ of two integers $\epsilon$ and $\kappa$. In the context of the spinless K-G equation, $\epsilon=0$ and $\kappa=l$ where $l$ is the orbital angular momentum quantum number. It is shown that both the eigenfunctions and energy eigenvalues of the K-G equation can also be expressed in terms of the $\eta_{0l}$-coupling function. The $\eta_{0l}$-coupling is therefore observed to completely encapsulate the concept of orbital angular momentum in the K-G formalism.

In section 3, it is found the $\eta_{\epsilon \kappa}$ function also pervades Dirac theory except that in this case $\epsilon=1$ and $\kappa=\pm(j+\frac{1}{2}) $ where $j$ is the total angular momentum quantum number of the electron including both orbital angular momentum and spin. On the basis of this result, the replacement $\eta_{0 l} \rightarrow \eta_{1 \kappa}$ is considered as a means of including electron spin in the K-G equation. The spin-corrected K-G equation generates scalar eigenfunctions and contains no matrices. It is solved here to generate  the same energy eigenvalues of hydrogen atom as the Dirac equation. 

\section{The Spin-0 K-G Equation}
The K-G equation determining the wavefunction $\psi$ for an electron of mass $m_0$ at a radial distance $r$ from a proton can be expressed in the form
\begin{equation} \label{eq: KGHA0}
-c^2 \nabla^2 \psi + \frac{m_0^2c^4}{\hbar^2}\psi = \left( \imath\frac{\partial}{\partial t} + c \frac{\alpha}{r} \right)^2 \psi
\end{equation}
where
\begin{equation} \label{eq: totE}
E\psi = \imath \hbar \frac{\partial \psi}{\partial t}
\end{equation}
is the total energy of the electron, $\alpha$ is the fine structure constant, $c$ is the velocity of light and $\hbar$ is Planck's constant divided by $2\pi$. Eqs. (\ref{eq: KGHA0}) and (\ref{eq: totE}) thus constitute an approximate model of the hydrogen atom neglecting the spin of the electron and the finite mass of the proton. 

The solution \cite{JN} to eqs. (\ref{eq: KGHA0}) and (\ref{eq: totE}) in spherical polar coordinates $(r,\theta,\phi)$ takes the separable form
\begin{eqnarray} \label{eq: psi_ha} 
\psi_{\epsilon nlm}(r,\theta,\phi,t) = R_{\epsilon nl}(r)Y_{lm}(\theta, \phi)\exp(-\imath E_{\epsilon nl}t / \hbar)
\end{eqnarray}
where $n,l,m$ are the hydrogen quantum numbers. The additional index $\epsilon$ has been included so the results in this section can be part of a more general discussion in later sections. For the purposes of this section, $\epsilon$ can be set equal to zero. Eq. (\ref{eq: KGHA0}) separates to give the radial equation
\begin{equation} \label{eq: KGHA0_R1}
\left[ \frac{1}{r^2}\frac{\partial}{\partial r} \left( r^2 \frac{\partial}{\partial r} \right) + \frac{E_{0nl}^2}{\hbar^2 c^2} + \frac{2E_{0nl}}{\hbar c}\frac{\alpha}{r} - \frac{m_0^2c^2}{\hbar^2} + \frac{\alpha^2}{r^2} - \frac{l(l+1)}{r^2}\right]R_{0nl} = 0
\end{equation} 
alongside the orbital angular momentum equation
\begin{equation}
\hat{L}^2Y_{lm} = l(l+1)Y_{lm}
\end{equation} 
where $\hat{L}$ is the orbital angular momentum operator.

The solution of eq. (\ref{eq: KGHA0_R1}) is known to take the form
\begin{equation} \label{eq: rwav_kg0}
R_{0nl}(r) = \frac{\mathcal{N}_{0nl}}{r^{\eta_{0 l}}} \exp \left( -\frac{r}{r_{0nl}} \right)\sum_{k=0}^{n} a_k r^k
\end{equation}
where 
\begin{equation} \label{eq: couplingFunc}
\eta_{\epsilon \kappa} = \frac{1+\epsilon}{2} \pm \sqrt{\left(\kappa+\frac{1-\epsilon}{2} \right)^2-\alpha^2}
\end{equation}
$\mathcal{N}_{0nl}$, $r_{0nl}$ and $a_k$ are constants. It is helpful that eq. (\ref{eq: couplingFunc}) can also be used to simplify eq. (\ref{eq: KGHA0_R1}) to give
\begin{equation} \label{eq: KGHA0_R2}
\left[ \frac{1}{r^2}\frac{\partial}{\partial r} \left( r^2 \frac{\partial}{\partial r} \right) + \frac{E_{0nl}^2}{\hbar^2 c^2} + \frac{2E_{0nl}}{\hbar c}\frac{\alpha}{r} - \frac{m_0^2c^2}{\hbar^2} + \frac{\eta_{0 l}(1-\eta_{0 l})}{r^2} \right]R_{0nl} = 0
\end{equation} 
Inserting the wavefunction (\ref{eq: rwav_kg0}) into eq. (\ref{eq: KGHA0_R2}) it is readily shown that 
\begin{eqnarray} \label{sum1}
\sum_{k=0}^n a_k  \left\{ \frac{k(k-1)r^{k-2}}{r_{0nl}^{2}} + \left[(1-\eta_{0 l})\frac{r_{0nl}}{r}-1 \right]\frac{2kr^{k-1}}{r_{0nl}^{2}} \right\}
+ \nonumber \\ 
\left[\frac{E_{0nl}^2}{\hbar^2 c^2} + \frac{2E_{0nl}}{\hbar c}\frac{\alpha}{r} - \frac{m_0^2c^2}{\hbar^2} - \frac{2(1-\eta_{0l})}{r_{0nl}r}+\frac{1}{r_{0nl}^{2}}\right]\sum_{k=0}^n a_k r^k  = 0 
\end{eqnarray} 
In this, terms in $r^n$ and $r^{n-1}$ equate to give
\begin{equation} \label{bohrRadius}
r_{0nl} = \frac{\hbar^2c^2}{\sqrt{E^2_{0nl} - m_0^2c^4}}
\end{equation} 
\begin{equation}
E_{0nl} = \frac{\hbar c (n+1-\eta_{0l})}{\alpha r_{0nl}}
\end{equation} 
Combining these two results leads to the energy eigenvalues for the hydrogen atom from the spinless K-G equation:
\begin{equation} \label{energy_ha}
E_{0nl} = m_0c^2 \left[ 1 + \frac{\alpha^2}{(n+1-\eta_{0l})^2}\right]^{-1/2}
\end{equation}
Note, eq (\ref{eq: couplingFunc}) for $\eta_{\epsilon k}$ contains a $\pm$ sign. The negative sign is usually chosen since in this case eq. (\ref{energy_ha}) corresponds to the Sommerfeld energy spectrum for the hydrogen atom. By comparison, the positive sign predicts a much higher binding energy sometimes called the hydrino state.

Clearly, the $\eta_{0l}$-function encapsulates the orbital angular momentum quantum number $l$ in eqs. (\ref{eq: KGHA0_R2}) through (\ref{energy_ha}). It can therefore be said that that angular momentum influences energy-momentum in the spin-0 K-G equation through the $\eta_{0l}$-coupling function.   

\section{The Spin-$\frac{1}{2}$ K-G Equation}
The Dirac equation for the hydrogen atom has two 4-component solutions of the form
\begin{eqnarray} 
\Psi^{\pm}=\left( \begin{array}{r}
\pm v(r)\chi^{\pm}(\theta,\phi) \\
u(r)\chi^{\mp}(\theta,\phi) \\
\end{array} \right)
\end{eqnarray} 
where $\chi^{\mp}(\theta,\phi)$ are spinors,
\begin{equation} \label{eq: dirac_u}
u(r) = r^{-\eta_{1 \kappa}} \exp \left( -\frac{r}{r_{1N\kappa}} \right) \sum_{k=0}^{N} a_k r^k
\end{equation},
\begin{equation} \label{eq: dirac_v}
v(r) = r^{-\eta_{1 \kappa}} \exp \left( -\frac{r}{r_{1N\kappa}} \right) \sum_{k=0}^{N} b_k r^k
\end{equation}
are scalar functions and
\begin{equation} \label{bohrRadiusD} 
r_{1N\kappa} = \frac{\hbar^2c^2}{\sqrt{E^2_{1N\kappa} - m_0^2c^4}}
\end{equation} 
Here, $E_{1N\kappa}$ denotes the energy eigenvalues describing the fine structure of the hydrogen atom. For a more detailed comparison to the literature \cite{DFL}, the integer $\kappa$ is related to the total angular momentum quantum number $j$ through the expression
\begin{equation}
\kappa = \pm\left(j+\frac{1}{2} \right)
\end{equation}
and
\begin{equation} \label{bigN} 
N=n-j-\frac{1}{2}
\end{equation}
where n is the principle quantum number of the hydrogen atom.

It is clear from eqs. (\ref{eq: dirac_u}) and (\ref{eq: dirac_v}) that the $\eta_{1\kappa}$-coupling function encapsulates total angular momentum in the Dirac formalism as the $\eta_{0 l}$-coupling function encapsulates orbital angular momentum in the spinless K-G equation. It is of interest on the strength of this result, to consider the replacement
\begin{equation}
\epsilon = 1, \quad n \rightarrow N, \quad l \rightarrow \kappa
\end{equation}
as a possible means of introducing electron spin into the Hamiltonian of the K-G equation (\ref{eq: KGHA0_R2}). This gives
\begin{equation} \label{eq: KGHA1_R1}
\left[ \frac{1}{r^2}\frac{\partial}{\partial r} \left( r^2 \frac{\partial}{\partial r} \right) + \frac{E_{1N\kappa}^2}{\hbar^2 c^2} + \frac{2E_{1N\kappa}}{\hbar c}\frac{\alpha}{r} - \frac{m_0^2c^2}{\hbar^2} + \frac{\eta_{1 \kappa}(1-\eta_{1 
\kappa})}{r^2} \right]R_{1N\kappa} = 0
\end{equation} 
to be a trial form of the radial K-G equation for spin-$\frac{1}{2}$ particles. 

Eqs. (\ref{eq: KGHA0_R2}) and (\ref{eq: KGHA1_R1}) are identical in form. The solution to eq. (\ref{eq: KGHA1_R1}) is therefore readily inferred to be
\begin{equation} \label{eq: rwav_kg1}
R_{1N\kappa}(r)
 = \frac{\mathcal{N}_{1N\kappa}}{r^{\eta_{1 \kappa}}} \exp \left( -\frac{r}{r_{1N\kappa}} \right)\sum_{k=0}^{N} a_k r^k
\end{equation}
identical in form to eq. (\ref{eq: rwav_kg0}). Inserting this back into eq. (\ref{eq: KGHA1_R1}) gives
\begin{eqnarray} \label{sum2}
\sum_{k=0}^n a_k  \left\{ \frac{k(k-1)r^{k-2}}{r_{1N\kappa}^{2}} + \left[(1-\eta_{1 \kappa})\frac{r_{1N\kappa}}{r}-1 \right]\frac{2kr^{k-1}}{r_{1N\kappa}^{2}} \right\}
+ \nonumber \\ 
\left[\frac{E_{1N\kappa}^2}{\hbar^2 c^2} + \frac{2E_{1N\kappa}}{\hbar c}\frac{\alpha}{r} - \frac{m_0^2c^2}{\hbar^2} - \frac{2(1-\eta_{1 \kappa})}{r_{1N\kappa}r}+\frac{1}{r_{1N\kappa}^{2}}\right]\sum_{k=0}^n a_k r^k  = 0 
\end{eqnarray}
identical in form to eq. (\ref{sum1}).

In eq. (\ref{sum2}), terms in $r^n$ and $r^{n-1}$ equate to give eq. (\ref{bohrRadiusD}) and the relationship
\begin{equation} \label{energy_kg1}
E_{1N\kappa} = \frac{\hbar c (n+1-\eta_{1\kappa})}{\alpha r_{1N\kappa}}
\end{equation} 
respectively. Combining eqs. (\ref{bohrRadiusD} ) and (\ref{energy_kg1}) together leads to the expression:
\begin{equation} \label{energy_dirac1}
E_{1N\kappa} = m_0c^2 \left[ 1 + \frac{\alpha^2}{(N+1-\eta_{1\kappa})^2}\right]^{-1/2}
\end{equation}
These are the energy eigenvalues of the hydrogen atom based on the spin-$1/2$ K-G equation. Using eqs. (\ref{eq: couplingFunc}) and (\ref{bigN}), this result can be rewritten
\begin{equation} \label{energy_dirac2}
E_{1N\kappa} = m_0c^2 \left[ 1 + \frac{\alpha^2}{(n-|\kappa|+\sqrt{\kappa^2-\alpha^2})^2}\right]^{-1/2}
\end{equation}
identical to the energy eigenvalues from the Dirac equation for the hydrogen atom.

\section{Concluding Remarks}
It has been shown for both the spinless K-G and Dirac equations for the hydrogen atom that a coupling function exists to take account of the influence of angular momentum on the Hamiltonian for the electron. It has been further shown that if the coupling function for the spinning electron from the Dirac equation is used to replace the coupling function in the spinless K-G equation, the spin-corrected K-G equation can be solved exactly to give the same energy eigenvalues as the Dirac equation.

\end{document}